\documentclass[%
 aip,
 amsmath,amssymb,
 reprint,%
]{revtex4-1}

\usepackage{graphicx}
\usepackage{dcolumn}
\usepackage{bm}

\usepackage[utf8]{inputenc}
\usepackage[T1]{fontenc}
\usepackage{mathptmx}
\usepackage{etoolbox}
\usepackage{xcolor}

\newcommand{\com}[1]{}

\makeatletter
\def\@email#1#2{%
 \endgroup
 \patchcmd{\titleblock@produce}
  {\frontmatter@RRAPformat}
  {\frontmatter@RRAPformat{\produce@RRAP{*#1\href{mailto:#2}{#2}}}\frontmatter@RRAPformat}
  {}{}
}%
\makeatother
\begin{document}

\preprint{AIP/123-QED}

\title[Protected Solid-State Qubits]{Protected Solid-State Qubits}

\author{Jeroen Danon\textsuperscript{*}}
\affiliation{Center for Quantum Spintronics, Department of Physics, Norwegian University of Science and Technology, NO-7491 Trondheim, Norway}
\email{jeroen.danon@ntnu.no}

\author{Anasua~Chatterjee}
\affiliation{Center for Quantum Devices, Niels Bohr Institute, University of Copenhagen, 2100 Copenhagen, Denmark}

\author{Andr\'as Gyenis}
\address{Department of Electrical, Computer \& Energy Engineering, University of Colorado Boulder, Boulder, Colorado 80309, USA}

\author{Ferdinand~Kuemmeth}
\affiliation{Center for Quantum Devices, Niels Bohr Institute, University of Copenhagen, 2100 Copenhagen, Denmark}

\date{\today}

\begin{abstract}
The implementation of large-scale fault-tolerant quantum computers calls for the integration of millions of physical qubits with very low error rates. This outstanding engineering challenge may benefit from emerging qubits that are protected from dominating noise sources in the qubits' environment. In addition to different noise reduction techniques, protective approaches typically encode qubits in global or local decoherence-free subspaces, or in dynamical sweet spots of driven systems. We exemplify such protected qubits by reviewing the state-of-art in protected solid-state qubits based on semiconductors, superconductors, and hybrid devices. 
\end{abstract}

\maketitle

\section{\label{sec:level1}Protection in Qubit Systems}

The extraordinary power of quantum computation stems from the exponentially large amount of information that can be associated with the entanglement in multi-qubit systems.\cite{nielsen2010quantum}
This implies that quantum information is in principle only useful if it can stay perfectly coherent for the whole duration of a computation.
However, every physical qubit is embedded in a noisy environment and the required access to qubit control unavoidably couples the qubit to this environment.
The coupling can result in relaxation and dephasing, and the resulting loss of quantum coherence in the system directly translates to errors in the quantum information.

Since decoherence is a generic problem, effective error mitigation and correction are quintessential for the development of reliable quantum computation in practice.\cite{Lidar2013}
Soon after the first quantum computation algorithms, powerful algorithms for error correction were proposed,\cite{Shor95,Steane96,Knill01} typically relying on redundancy in the encoding and the measurement of syndrome operators followed by active correction of the errors.
These active methods can require a large overhead, both in terms of required information storage capacity and complexity of the actual algorithms.
For this reason, an emerging activity is on qubits with some kind of \textit{passive} error resilience, often built into the system Hamiltonian.

\begin{figure}[t]
\includegraphics[width=\columnwidth]{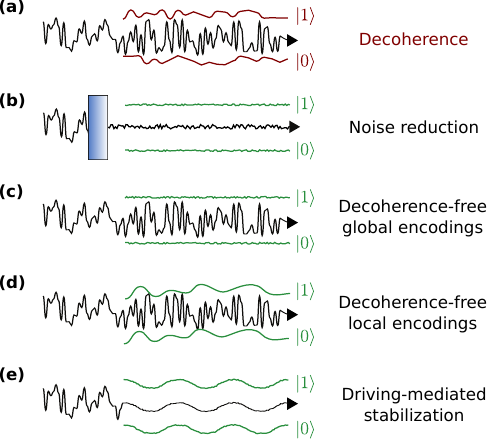}
\caption{(a) Local fluctuations in the environment can couple to the qubit subsystem, leading to qubit dephasing and relaxation. Several approaches can yield a protected qubit. (b) The harmful fluctuations can be removed or filtered out. (c) The qubit can be encoded into global degrees of freedom that are insensitive to local fluctuations. (d) The qubit can be smartly encoded into a subsystem of local degrees of freedom that couple in such a symmetric way to the environment that the fluctuations do not cause any uncontrollable qubit dynamics. (e) Strong external driving can in some cases stabilize the qubit dynamics.}
\label{fig1}
\end{figure}

We classify the efforts in this direction into four categories (as illustrated in Fig.~\ref{fig1}):

(i) The most straightforward strategy for protecting a qubit is to identify the dominating source of noise contributing to decoherence and then tailor the design and the environment of the qubit to remove or reduce that particular noise before it reaches the qubit [Fig.~\ref{fig1}(b)].
This \textit{noise-filtering} approach is very effective, but not always easy to realize in practice.
    
(ii) Since environmental fluctuations are typically local in nature, one can encode qubits into \emph{global} degrees of freedom of the host system that are insensitive to local fluctuations [Fig.~\ref{fig1}(c)].
This is the main idea behind topological qubits,\cite{Stanescu17} where the qubit control mechanisms couple in a topologically different way to the system than the noise sources.

\begin{figure*}[t]
\includegraphics[width=\textwidth]{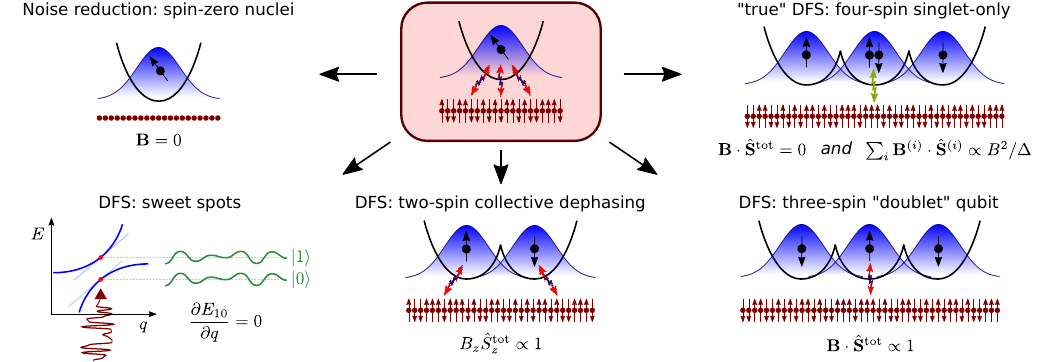}
\caption{Different strategies to mitigate magnetic noise in semiconductor spin qubits. Local DFS-based approaches rely on multi-spin encodings, which often come with significant spin-charge hybridization, making the qubit susceptible to charge noise. Tuning the qubit to a sweet spot, where the two states couple to linear order in the same way to the (charge) noise results in a protection against dephasing similar to the mechanism of collective dephasing in two-spin encoded qubits.
}
\label{fig2}
\end{figure*}

(iii) One can encode the qubits into local degrees of freedom, but use a subspace of the Hilbert space that couples to the fluctuations of concern such that the environment cannot distinguish the different qubit states [Fig.~\ref{fig1}(d)].
These subspaces are commonly referred to as \emph{decoherence-free subspaces},\cite{Lidar98} and can offer various degrees of protection depending on the complexity of the encoding.
In general, this type of protection works best for long-wavelength noise, when fluctuations are uniform on the physical scale corresponding to the subspace used for the encoding.

(iv) Encoding qubits into time-dependent states is another emergent approach towards increasing coherence [Fig.~\ref{fig1}(e)]. In this approach, the qubits are subject to strong external driving fields that can compete with the noise and stabilize the qubit states, for example through autonomous error correction\cite{gertler_protecting_2021} or by creating dynamical sweet spots.\cite{ziwen2020,mundada2020,didier}

In this article, we discuss the progress and current state of the art in the implementation of protected qubit systems in three solid-state platforms: semiconductor spin qubits,  superconducting qubits and Majorana zero-mode-based qubits.

\section{Protected semiconductor spin qubits}

Due to their small size and fast operation times, gate-controlled semiconductor spin qubits are considered an attractive candidate platform for massively scaling quantum information processing.\cite{Chatterjee2021}
A primary source of decoherence for the first generation of spin qubits was the hyperfine interaction between the localized electron spins and the randomly fluctuating nuclear spins of the semiconductor, effectively leading to magnetic noise that causes unwanted qubit rotations.\cite{Hanson2007}
So far, the most effective mitigation strategy for this noise has been one of \emph{noise reduction}:
Using group-IV semiconductors such as Ge and Si, instead of GaAs or InAs, allows to work with isotopically purified nuclear-spin-free samples,\cite{Zwanenburg2013,Scappucci2020} which resulted in an extension of coherence times by more than four orders of magnitude.\cite{Stano2021}
Apart from this effective and conceptually simple, though technically challenging approach, also more sophisticated (yet similarly passive) spin-qubit protection strategies have been proposed and implemented over the years, a few of which we will highlight below, see also Fig.~\ref{fig2}.

As one attractive feature of spin qubits is their small size, protection schemes based on \emph{local} encodings in decoherence-free subspaces (DFSs) are particularly interesting, because these encodings do not require a qualitative scale-up in qubit size, as opposed to global (e.g., topological) encodings.

A local DFS is often defined in a subspace of the Hilbert space spanned by a few qubits.\cite{Lidar2003}
The simplest multi-qubit (in this case multi-spin) encoding that provides some built-in protection against magnetic noise is the singlet-triplet qubit, which is defined in the $S_z^{\rm tot}=0$ subspace of two spins localized in two quantum dots.\cite{Levy2002,Petta2005}
The two states in this subspace couple in exactly the same way to a magnetic field that is uniform on the scale of the double quantum dot, and the subspace is thus insensitive to (long-wavelength) magnetic noise along the quantization axis $z$.
This type of protection is more generally known as \emph{collective dephasing}\cite{Palma1996,Duan1998} and boils down to the following observation. If two qubits suffer the same pure dephasing, i.e., $\alpha |{0}\rangle + \beta|{1}\rangle \to \alpha |{0}\rangle + e^{i\varphi}\beta|{1}\rangle$, then any state in the two-dimensional space spanned by $\{|{10}\rangle,|{01}\rangle\}$ only acquires an unimportant overall phase factor. 
However, since the nuclear magnetic noise is short-wavelength (not strongly correlated across different qubits) and not uniaxial, it does not result in collective dephasing and the two-spin encoding does not provide significant protection against it.

\begin{figure*}[t]
\includegraphics[width=\textwidth]{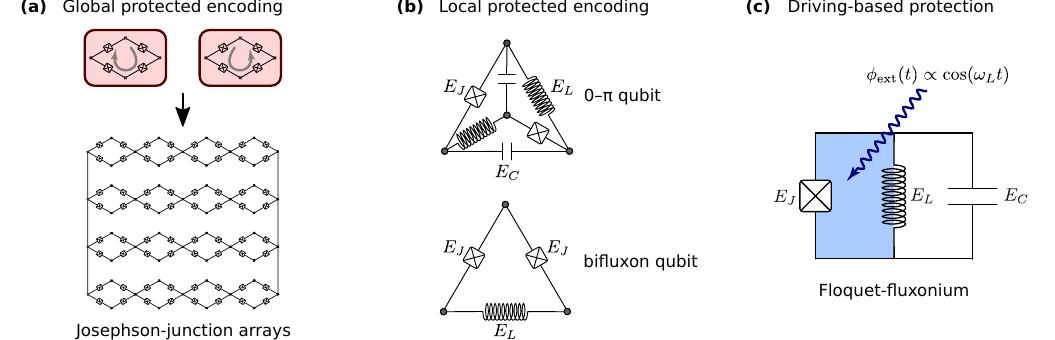}
\caption{Examples of protected superconducting qubits. (a) The rhombi superconducting circuit (top), which consists of four identical Josephson junctions in a single loop, can encode a qubit into quasi-degenerate quantum states, but these states are extremely sensitive to flux noise. By building an array of unprotected qubits one can increase protection against dephasing due to flux noise (bottom). (b) The $0$--$\pi$ circuit and the bifluxon qubit are two examples of locally encoded protected superconducting circuits with only a few degrees of freedom. (c) The Floquet-fluxonium qubit is a qubit whose protection is the result of periodically modulating the flux through the loop of the device, which creates dynamical sweet spots.}
\label{fig3}
\end{figure*}

It is worth pointing out that collective dephasing is conceptually similar to the mechanism that underlies improved qubit coherence at so-called \emph{sweet spots}.
In multi-dot multi-spin qubits the interdot tunnel coupling leads to spin-charge hybridization.
While this allows fast electrical control of the qubits, it makes the qubit sensitive to ubiquitous fluctuations of the electric field, coming from the substrate and/or the gate electrodes.
One approach to mitigate such charge noise, without compromising the electric control over the qubit, is to operate the qubit at sweet spots, where the qubit splitting $E_{10}$ is to lowest order insensitive to the dominating electric fluctuations:
Small fluctuations then translate to an irrelevant overall phase factor for the qubit subspace.
The sweet-spot approach simply amounts to choosing optimal working points in parameter space and has proven to be capable of significantly increasing the charge-noise-resilience of multi-spin qubits.\cite{Martins2016,Reed2016}
In this context, one can say that mitigating noise through collective dephasing is an instance of the same approach: It results in a subspace with a special, prolonged sweet spot against the particular fluctuations causing the dephasing.

A more advanced local DFS, providing protection against magnetic noise along all three axes, can be formed by combining three spins:\cite{Knill1999,Filippo2000,Yang2001}
A three-spin system hosts one $S=\frac{3}{2}$ spin-quadruplet and two $S=\frac{1}{2}$ spin-doublets.
A qubit encoded into two \emph{identical} spin-$\frac{1}{2}$ states but belonging to \emph{different} doublets evolves trivially under the Hamiltonian $\hat H = {\bf B}(t)\cdot\hat{\bf S}^{\rm tot}$ (with $\hat{\bf S}^{\rm tot}$ the total spin operators) and is thus protected against all long-wavelength magnetic noise.
This protection is better than that of the two-spin qubit, but still inefficient against short-wavelength nuclear magnetic noise.

Nevertheless, three-spin qubits implemented in triple quantum dots have drawn much attention for a different reason:
The two independently controllable exchange couplings between neighboring dots translate into \emph{two} electrically tunable control axes on the qubit's Bloch sphere.\cite{DiVincenzo2000,Laird2010,Medford2013,Medford2013a,Russ2017}
This so-called exchange-only qubit thus offers all-electric control, at the price of an increased sensitivity to charge noise.
As for the two-spin case, operation of this qubit at a (multiple) sweet spot can efficiently reduce the effects of charge noise.\cite{Fei2015,Russ2016,Shim2016,Zhang2017}.

Finally, using four qubits one can create a ``true'' local DFS,\cite{Lidar98} which is probably most easily understood in terms of spin:
The four-spin Hilbert space contains \emph{two} $S=0$ spin-singlet subspaces, and a qubit encoded into the two corresponding spinless states does not couple to any uniform magnetic field.
The important advantage over the three-spin encoding is that the singlet-only qubit also provides some protection against \emph{local} magnetic fluctuations:\cite{Lidar98}
The effect of these fluctuations becomes higher-order and is suppressed by a factor $B/\Delta$, where $B$ characterizes the magnitude of the fluctuating fields and $\Delta$ the typical energy splitting between the singlets and the other states.
Recently, several quantum-dot-based implementations of the singlet-only spin qubit were proposed,\cite{Sala2017,Russ2018,Sala2020} all of which offer the same exchange-based all-electric control as the three-spin exchange-only qubit.

\section{Protected superconducting qubits}

Since the first realizations of superconducting qubits, the idea to create quantum states that are robust against environmental noise has fueled intense research activity.\cite{kjaergaard2020, gyenis2021b} Although most superconductor-based quantum devices are constructed from only three basic circuit elements, (capacitors, inductors and aluminum-oxide-based Josephson junctions), the combination of these building blocks offers outstanding flexibility. For example, the number of nodes in the circuit determines the dimension of the quantum system, the size of inductors and Josephson junctions fixes the potential, and the capacitance values define the kinetic energy of the quantum state.\cite{vool2017} The ability to independently control these three aspects of a quantum state (dimension, kinetic and potential energy) allows to construct superconducting circuits that satisfy the requirements for intrinsic noise protection.
Moreover, it has been shown theoretically that some of these qubits allow protected gate operations as well,\cite{brooks2013,Klots2021} another crucial ingredient toward fault-tolerant quantum computation.

The development and design of protected superconducting qubits have evolved along three main paths: \textit{Josephson-junction-array-based multimode circuits},\cite{kitaev2003,ioffe2002a,ioffe2002b,doucot2002,doucot2005,gladchenko2008,Doucot2012,bell2014} \textit{compact few-mode circuits},\cite{brooks2013,smith2020,Kalashnikov2020} and \textit{driven systems}\cite{ofek2016,mirrahimi2014,puri2017,campagne2020,grimm2020,ziwen2020,mundada2020,didier} (see Fig.~\ref{fig3}).
In the first approach, the logical qubit is constructed by concatenating a set of noisy circuit unit cells to reduce the total effect of local errors along the chain (a type of global encoding). This approach to noise protection is rooted in the concept of a topological ground state degeneracy. On the other hand, compact protected circuits are individual qubits with only a few degrees of freedom, and are examples of qubits in local decoherence-free subspaces. Generally, to achieve noise protection, these circuits need to meet extreme requirements, such as large inductors and a strong reduction of parasitic capacitances. Protection in these compact qubits often arises because the low-energy behavior of the circuit can be approximated by an effective Hamiltonian with parity symmetry such as Cooper-pair or fluxon symmetry. Finally, systems under intense continuous driving can also show protection against various noise sources. In this case, the rich interaction of the driving field and the qubit leads to advantageous properties. Below we discuss examples for each of these approaches. 

We first discuss the case of the Josephson-junction-array qubits [Fig.~\ref{fig3}(a)]. The basic building block of such arrays is often referred to as the “$\cos2\phi$” circuit. There are multiple ways to realize these $\cos2\phi$ circuits \cite{doucot2002,gladchenko2008,bell2014,larsen2020,smith2020}, but a common feature is that they have potential landscapes with two minima as a function of a phase bias $\phi$ of the superconductor order parameter (effectively a double-well potential with $\pi$ periodicity) and discrete charge states. The qubit states are quasi-degenerate and can be described by an effective Cooper-pair parity: the ground state wavefunction contains an even number of Cooper pair states, while the excited state has an odd number. Such a $\cos2\phi$ element, in principle, could be a protected qubit by itself: transitions between the qubit states are not allowed—as long as the noise does not break the parity symmetry—and dephasing is suppressed because the qubit energy is insensitive to external fluctuations (degenerate levels). But it is important to highlight that the realized circuits behave as $\cos2\phi$ elements only at a very narrow range of external parameters and when the circuit elements are highly symmetric.\cite{doucot2002,gladchenko2008,bell2014,larsen2020,smith2020} For example, flux noise can drive the qubit out of the $\cos2\phi$ regime, easily destroying the protection. The idea behind the concatenation is to combat the flux noise: when multiple of these units are connected to each other, the effect of flux noise is reduced, and the chain is less sensitive to the parity breaking effects of local flux noise.\cite{Doucot2012} As a result, the logical qubit is more robust, and its sensitivity to flux fluctuations is suppressed approximately as an exponential function of the number of unit cells used in the chain.

Regarding the second approach,\cite{gyenis2021b} the prototypical examples for compact protected qubits are the $0$--$\pi$ qubit\cite{kitaev2006,brooks2013,dempster2014,groszkowski2018,dipaolo2019,gyenis2019} and the bifluxon qubit\cite{Kalashnikov2020} [Fig.~\ref{fig3}(b)]. Focusing on the $0$--$\pi$ qubit here, this circuit has only three degrees of freedom, and when the qubit parameters meet strict requirements, the Hamiltonian of the device can be approximated with an effective $\cos2\phi$ Hamiltonian. In contrast to the noisy $\cos2\phi$ elements discussed above, the $0$--$\pi$ qubit behaves as a $\cos2\phi$ qubit at all external parameters and it has exponentially reduced sensitivity to noise. Thus, there is no need to concatenate multiple of these qubits to secure the protection: the qubit is protected by itself. The challenge of realizing this qubit is that the requirements for the circuit parameters is beyond the currently available materials and fabrication procedures. In the experimentally realized soft-$0$--$\pi$ qubit version of the device, the protection against flux noise is reduced from exponential protection to first-order sweet-spot protection.\cite{gyenis2019}

The third common approach towards noise protection in superconducting qubits is to expose the quantum circuit to periodically modulated external fields [see Fig.~\ref{fig3}(c)]. For example, in a recent work, the flux-sensitivity of a superconducting qubit, the fluxonium, was significantly reduced by strongly modulating the external flux at frequencies close to the qubit transition frequency.\cite{didier,didier2,ziwen2020, mundada2020} The interplay of the field and the qubit levels leads to dynamical sweet spots in certain parameter regimes of the drive, where the Floquet-fluxonium qubit becomes first-order protected against flux noise.

\begin{figure}[t]
\includegraphics[width=\columnwidth]{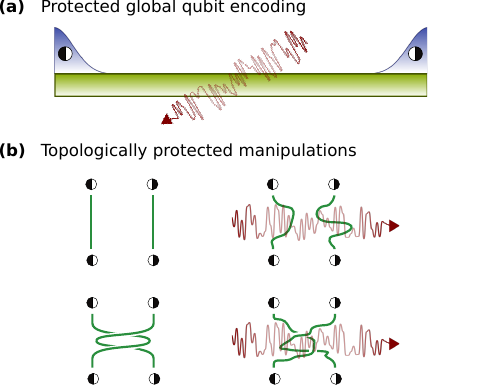}
\caption{Topological quantum computation: (a) Qubits are encoded globally in low-energy modes resulting in a decoupling from local fluctuations. (b) Some qubit gates could be performed by braiding the localized modes, the outcome of which is independent of the local details of the trajectories followed by the modes.
}
\label{fig4}
\end{figure}

\section{Topologically protected qubits}

In recent years topological qubit encodings, as mentioned under point (ii) in the introduction, gained much attention and topology became almost a synonym for protection. The idea that hybrid semiconductor-superconductor devices, based on new materials and fabrication techniques, could be tuned to a topologically non-trivial phase that could be used for protected qubit encodings\cite{Sato2009,Sau2010,Alicea2010} fueled an explosion of research in this direction, both theoretical and experimental.

Most notably, a quasi-one-dimensional semiconductor with strong spin–orbit coupling and a large Landé g-factor could, when proximitized by an $s$-wave superconductor, effectively become a topological superconductor with exponentially localized Majorana zero-modes (MZMs) at its ends.\cite{Oreg2010,Lutchyn2010,Hell2017}
These MZMs are decoupled from local fluctuations and are predicted to obey non-Abelian anyonic exchange statistics.
Therefore, if successful, qubits encoded into a space spanned by such MZMs would not only have a decoherence that is exponentially suppressed in the physical separation between the MZMs [as illustrated in Fig.~\ref{fig4}(a)], but they would also allow to perform topologically protected gate operations through braiding [Fig.~\ref{fig4}(b)], although, due to the Ising-like nature of the anyons, the resulting set of topologically protected quantum gates is not universal.\cite{Stanescu17}

The built-in topological qubit protection is indeed very appealing, justifying the intense research efforts in this direction. However, so far a clear experimental demonstration of the existence of non-Abelian low-energy modes in solid-state devices is still lacking.
One of the barriers that have emerged is the materials-science challenge of realizing effective spinless $p$-wave superconductivity\cite{lutchyn2018}. Experiments so far have focused on nanowires and 1D-systems with strong spin–orbit interaction, proximity-coupled to an $s$-wave superconductor, typically taking the form of a heavy-element semiconductor such as InAs and InSb, coupled to Al and NbTiN as the superconducting components. However, disorder at the semiconductor–superconductor interface is likely the source of the so-called soft gap, or continuum of subgap states.\cite{Takei2013} These states, yielding a large subgap conductance, degrade the hard superconducting gap that protects against thermal quasiparticle excitations. This is challenging for the topological phase, which is stable with respect to small perturbations only as long as they do not cause the bulk gap to collapse.\cite{Akhmerov2011} Efforts to resolve this materials-science challenge have concentrated on interface improvement, with progress made in growing high-quality thin Al films on pristine nanowires,\cite{Krogstrup2015} without exposing the interface to air and oxidation, as well as eliminating scattering sites and disorder within the semiconductor itself.

A second, related challenge has been the difficulty of detecting unambiguous signatures of MZMs in a typical conductance experiment.
The textbook case resulting in non-local entanglement in gapped systems relies on an unpaired MZM residing on each end of the nanowire.\cite{kitaev2001}
However, if each end of the wire hosts an even number of MZMs, they can form a local and conventionally fermionic Andreev state lying near zero energy.
Partially separated Andreev states, more concerningly, separated by a distance of the order of the characteristic Majorana decay length, can have nearly zero energy over a large range of the Zeeman field, chemical potential, and tunnel barrier height, generating signatures identical to MZMs in local charge tunneling experiments (performed on one side of the nanowire).\cite{Moore2018}
Therefore, experiments need to distinguish trivial low-energy modes and MZMs, which is difficult using any type of local measurement at one nanowire end of the wire such as a charge or spin tunneling measurement.
Considerably more complex and technically demanding experiments, many of which are currently ongoing in research laboratories, are thus required to simply confirm the topologically nontrivial state, such as synchronized two-terminal charge tunneling measurements, quasiparticle interference, fusion, or braiding.
For a recent review of this fast-evolving field we refer the reader to Ref.~\onlinecite{lutchyn2018}.

\section{Outlook}

Despite significant recent progress, creating a full-scale quantum computer remains very challenging, not in the least because of the large ratio of physical to logical qubits needed for quantum error correction.
Qubits with inherent protection against specific errors, as discussed in this article, have the potential to strongly relax this requirement; a focus on the design and implementation of such qubits thus seems to be a promising strategy for realizing near-term breakthroughs in the functionality of quantum processors.
In this context, the overview of different possible approaches to qubit protection we outlined in the introduction, and exemplified in the following sections, could provide some guidance when seeking the optimal protection strategy for a given noisy qubit.

When designing protected qubits, their possible drawbacks as well as the required overhead for implementing quantum codes have to be kept in mind, for example:
(i) When enlarging the total Hilbert space used to host a qubit, e.g., to accommodate a decoherence-free subspace, the number of leakage channels out of the computational subspace may increase, leading to enhanced decoherence, or accessing the qubit subspace for initialization and readout may become more intricate and prone to errors.
(ii) Sweet spots where the qubit is insensitive to the dominating source of noise sometimes unavoidably come with smaller coupling to control pulses.
(iii) Two-qubit gates are more challenging due to the protected nature of the separate qubits.
All such potential difficulties need to be understood for each protected qubit implementation that is considered as a basis for fault-tolerant quantum computation.
In the future one may thus expect quantum processors based on a heterogeneous qubit system: quantum memories implemented in long-lived protected qubits (ideally with a low overhead of classical computing power), gate operations performed in more controllable qubits or using intrinsically protected gates, and reliable transfer of quantum information between different types of qubit.

\begin{acknowledgments}
JD acknowledges support via FRIPRO-project 274853, which is funded by the Research Council of Norway (RCN).
\end{acknowledgments}

\section*{Data Availability Statement}
Data sharing is not applicable to this article as no new data were created or analyzed in this
study.

\bibliography{Library}

\end{document}